# The HKV method of solving of replicator equations and models of biological populations and communities

G.P. Karev

National Center for Biotechnology Information, National Institute of Health,
Bldg. 38A, Rm. 5N511N, 8600 Rockville Pike, Bethesda, MD 20894, USA
E-mail: *karev@ncbi.nlm.nih.gov*

**Abstract.** Replicator equations (RE) are among the basic tools in mathematical theory of selection and evolution. The method of Hidden keystone variables (HKV) for reducing a wide class of RE of high or even infinite dimensionality to low-dimensional escort systems of ODEs, which can be explored analytically in many cases was developed. The method has potential for different applications to models of biological populations and communities

1. **Statement of the problem**

The selection system is a mathematical model of an inhomogeneous population, in which every individual is characterized by a vector-parameter $\mathbf{a} = (a_1, \ldots a_n)$ that takes on values from set $A$. The parameter $\mathbf{a}$ specifies an individual's inherited invariant properties and does not change with time; the set of all individuals with a given value of the vector-parameter $\mathbf{a}$ in the population is called $\mathbf{a}$-clone.

Let $l(t, \mathbf{a})$ be the density of the population at time $t$ over the parameter $\mathbf{a}$, so that the total population size $N(t) = \int_A l(t, \mathbf{a}) d\mathbf{a}$ and the current population distribution $P(t, \mathbf{a}) = l(t, \mathbf{a})/N(t)$. Denote $F(t, \mathbf{a})$ the per capita reproduction rate at time $t$. It is assumed that the reproduction rate of every $\mathbf{a}$-clone does not depend on other clones but can depend on $\mathbf{a}$ as well as on some general population characteristics of the system ("regulators"), having the form $G(t) = \int_A g(\mathbf{a}) l(t, \mathbf{a}) d\mathbf{a}$, where $g(\mathbf{a})$ is an appropriate function. The total population size is the most important regulator, corresponding to $g(\mathbf{a}) \equiv 1$.

In many cases the reproduction rate may depend on mean values over the population distribution of the form $E_t[g] = \int_A g(\mathbf{a})P(t,\mathbf{a})d\mathbf{a}$. Let us emphasize that these mean values can be computed with the help of corresponding regulators according to the formula $E_t[g] = G(t)/N(t)$.

Overall, we specify for each model a finite set of regulators $\mathbf{G}(t) = \{G_1(t), \dots G_m(t)\}$, which contains the total population size; we assume that the individual reproduction rate can depend on this set of regulators at each time moment. Then the population dynamics can be described by the following master model of *selection system*:

$$\frac{dl(t,\mathbf{a})}{dt} = l(t,\mathbf{a})F(t,\mathbf{a}), \tag{1.1}$$

$$F(t,\mathbf{a}) = \sum_{i=1}^{n} u_i(t,\mathbf{G})\varphi_i(\mathbf{a}), \tag{1.2}$$

where $u_i(t,\mathbf{G})$ are continuous functions (see details and examples in [16],[17]). The initial distribution $P(0,\mathbf{a})$ and the initial population size $N(0)$ are assumed to be given.

The mathematical form of the reproduction rate suggests (from a biological point of view) that the individual's fitness depends on a given finite set of traits. The function $\varphi_i(\mathbf{a})$ in (1.1) may describe quantitative contribution of a particular *i*-th trait to the total fitness and then $u_i(t,\mathbf{G})$ describes the relative importance (weight) of the trait contributions, which at every time moment can depend on the state of the environment, population size, the mean, variance, covariance, and other statistical characteristics of the traits. The current probability distribution solves the *replicator equation*:

$$\frac{dP(t,\mathbf{a})}{dt} = P(t,\mathbf{a})(F(t,\mathbf{a}) - E_t[F]) \tag{1.3}$$

Replicator equations are among the basic tools in mathematical theory of selection and evolution, see, e.g., [6].

## 2. Reduction theorem

In model (1.1) the regulators and hence the reproduction rate $F(t,\mathbf{a})$ are not given explicitly but should be computed using the current pdf $P(t,\mathbf{a})$ at each time moment, so the model in the general case is a nonlinear equation of infinite dimensionality. Nevertheless, it can be reduced to a Cauchy problem for the *escort system* of ODE of dimensionality equal to the number of traits. To this end, introduce the *generating functional*:

$$\Phi(r;\boldsymbol{\lambda}) = \int_A r(\mathbf{a})\exp(\sum_{i=1}^{n} \lambda_i \varphi_i(\mathbf{a}))P(0,\mathbf{a})d\mathbf{a} \tag{2.1}$$

where $\boldsymbol{\lambda} = (\lambda_1, \dots \lambda_n)$ and $r(\mathbf{a})$ is a measurable function on $A$.

Define *auxiliary (keystone) variables* as a solution to the *escort system* of differential equations:

$$\frac{dq_i(t)}{dt} = u_i(t, \mathbf{G}^*(t)), \; q_i(0) = 0, i = 1, \ldots n, \tag{2.2}$$

where $\mathbf{G}^*(t) = \{G_1^*(t), \ldots G_m^*(t)\}$ and

$G_k^*(t) = N(0)\Phi(g_k, \mathbf{q}(t)), \; \mathbf{q}(t) = \{q_1(t), \ldots q_n(t)\}$.

Denote $K_t(\mathbf{a}) = \exp(\sum_{i=1}^n q_i(t)\varphi_i(\mathbf{a}))$.

**Reduction Theorem** [14,15]. *Let $0 < T \leq \infty$ be the maximal value of $t$ such that Cauchy problem* (2.2) *has a unique global solution$\{\mathbf{q}(t)\}$ at $t \in [0, T)$. Then the functions:*

$$l(t, \mathbf{a}) = l(0, \mathbf{a})K_t(\mathbf{a}) \tag{2.3}$$

$$G_k(t) = G_k^*(t) = N(0)\Phi(g_k, \mathbf{q}(t)), \tag{2.4}$$

*satisfy System* (1.1)-(1.2) *at $t \in [0, T)$.*

In particular, the total size of the population

$$N(t) = N(0)\Phi(1, \mathbf{q}(t)) = N(0)E_0[K_t]. \tag{2.5}$$

As a corollary, we obtain the central formula for the current distribution of the system:

$$P(t, \mathbf{a}) = P(0, \mathbf{a})K_t(\mathbf{a})/E_0[K_t]. \tag{2.6}$$

The reduction theorem implies a powerful HKV (Hidden Keystone Variables) method for solving selection systems (1.1)-(1.2) and corresponding replicator equations (1.3). Introducing auxiliary variables, which solve the escort system of non-autonomous differential equations (2.2) (see also (2.8) below), allows us to write down the solution of selection system and replicator equation and to find all its statistical characteristics of interest.

The general HKV method is simplified in an important case of the reproduction rate $F(t, \mathbf{a}) = \sum_{i=1}^n u_i(t, \mathbf{G})\varphi_i(\mathbf{a})$ with the regulators of the form $N(t), G_i(t) = N(t)E_t[\varphi_i]$ only. In this case we can use the moment generating function of the joint initial distribution of the variables $\varphi_i$,

$$M(\boldsymbol{\lambda}) = E^0[\exp(\sum_{i=1}^n \lambda_i \varphi_i)], \tag{2.7}$$

instead of general functional (2.1). The escort system reads

$$\frac{dq_i}{dt} = u_i(t, \mathbf{G}(t)), q_i(0) = 0, i = 1, \ldots n \tag{2.8}$$

where $G_i(t)$ are defined via formulas $N(t) = N(0)M_0(\mathbf{q}(t)), E_t[\varphi_i] = \partial_i \ln M_0(\mathbf{q}(t))$.

The solution to corresponding replicator equation

$P(t, \mathbf{a}) = P(0, \mathbf{a}) K_t(\mathbf{a}) / E_0[K_t]$ where $E_0[K_t] = M_0(\mathbf{q}(t))$.

The following examples demonstrate the HKV method at work.

### 3. Examples

#### 3.1. *Inhomogeneous Malthusian model*

The simplest inhomogeneous *Malthusian* model is of the form

$$\frac{dl(t,a)}{dt} = a\, l(t,a) . \tag{3.1}$$

The model describes a population of clones $l(t, a)$ each of which grows at its own rate $a$. The corresponding replicator equation reads $\frac{dP(t,a)}{dt} = P(t,a)(a - E_t[a])$. The escort equation is trivial in this case, $\frac{dq}{dt} = 1$, and the keystone variable is just time.

Let $M(\lambda) = \int_A \exp(\lambda a)\, P(0, a) da$ be the mgf of the initial distribution. Then the solution of (3.1) is given by the formulas $l(t, a) = l(0, a) \exp(at)$, $N(t) = N(0) M(t)$, and the solution to the replicator equation $P(t, a) = P(0.a) \exp(at) / M(t)$.

We can see that even the simplest replicator equation possesses a variety of solutions depending on the initial distribution. Recall that according to the theorem of S. Bernstein, a function $M(\lambda)$ is the mgf for some pdf if and only if it is absolutely monotone and $M(0) = 1$. So, the total size of inhomogeneous Malthusian population can change like an arbitrary absolutely monotone function $M(t)$ at corresponding initial distribution. Next, $\frac{dN}{dt} = N E_t[a]$ and $\frac{dE_t[a]}{dt} = Var^t[a] > 0$ (it is the simplest version of the Fisher Fundamental theorem of selection [3]); hence, any inhomogeneous Malthusian population increases *hyper-exponentially*. See [11], [16] for details and different examples.

#### 3.2. *Global demography*

The growth of the world population up to ~1990 was described with high accuracy by the hyperbolic law [4]:

$$N(t) = \frac{C}{T-t} \text{ with } \approx 2 * 10^{11} . \tag{3.2}$$

Formula (3.2) describes the hyper-exponential growth of humankind. This formula solves the quadratic growth model

$$\frac{dN}{dt} = N^2/C, \qquad (3.3)$$

in which the growth rate is proportional to the number of "pair contacts" in the *whole* of humankind and the individual reproduction coefficient is proportional to the total world population. This fact is difficult to interpret from the point of view of elementary processes and it seems evidently wrong even for small populations. In addition, equations (3.2), (3.3) predict a non-realistic *demographic explosion* at some point in time as we approach the year $T \cong 2025$. This means that $N(t) \to \infty$ when $t \to T$, and the same is true for the population growth rate and individual reproduction coefficient.

Let us note that, usually, Eqs. (3.2) and (3.3) are considered to be equivalent; this assertion was the basis for a new phenomenological theory of global demography developed in [8], [9]. In the latter theory, the underlying assumption of the humankind uniformity and the interpretation of the "quadratic law" (3.3) in terms of the information community hypothesis are, perhaps, unrealistic, especially when dealing with pre-historic human populations that were composed of non-interbreeding subgroups. Let us emphasize that, paradoxically (from the standpoint of the theory developed in [8]), the hyperbolic growth was observed in the past but has substantially slowed down during the last decades when most of the world population indeed became an "information community".

The mathematical part of the theory was based on some modification of the quadratic equation (3.3), which was considered as an exceptional phenomenon inhered to humankind only.

In my opinion, equation (3.3) cannot be the basis of any realistic demographic theory because it apparently makes no "biological" sense, at least, for large populations. Then, the problem arises: why the unacceptable quadratic equation (3.3) implies a good fit of empirical data (3.2)? The answer is: the hyperbola $N(t) = \frac{C}{T-t}$ is implied not only by the quadratic growth model but also by more plausible Malthusian inhomogeneous model (3.1) (see [11] for details).

Given that any real population is inhomogeneous, the simplest inhomogeneous Malthusian model is more acceptable as a starting point for global demography modeling then the quadratic growth model. The population increases in such a way that the distribution of the reproduction rate is *exponential* at every instant $t < T$ with the mean

$E^t[a] = 1/(T-t).$

The "demographic explosion" occurs at the moment $t = T$ when not only $N(T) = \infty$, but also $E_T[a] = \infty$ and $Var_T[a] = \infty$. It is a corollary of the obviously wrong assumption (incorporated implicitly into quadratic growth model) that the individual reproduction rate may take arbitrary large values with non-zero probability. Hence, a natural way to eliminate the unrealistic "demographic explosion" from the model is to take onto account that possible values of the reproduction rate should be limited. When the reproduction rate $a$ in model (3.2) is bounded, i.e. the rate has an exponential distribution truncated in the interval $[0, c)$ (specifically for real demographic data, $c \approx 0.114$ [11]) then one can show that

$$N(t) = N(0)\frac{T}{T-t}(1 - exp^{-c(T-t)}). \tag{3.4}$$

It easily follows from (3.4) that $N(t)$ is finite, even though indefinitely increasing, for all $t$, and is very close to the hyperbola for a long time (up to 1990 at corresponding values of coefficients, see Fig.1).

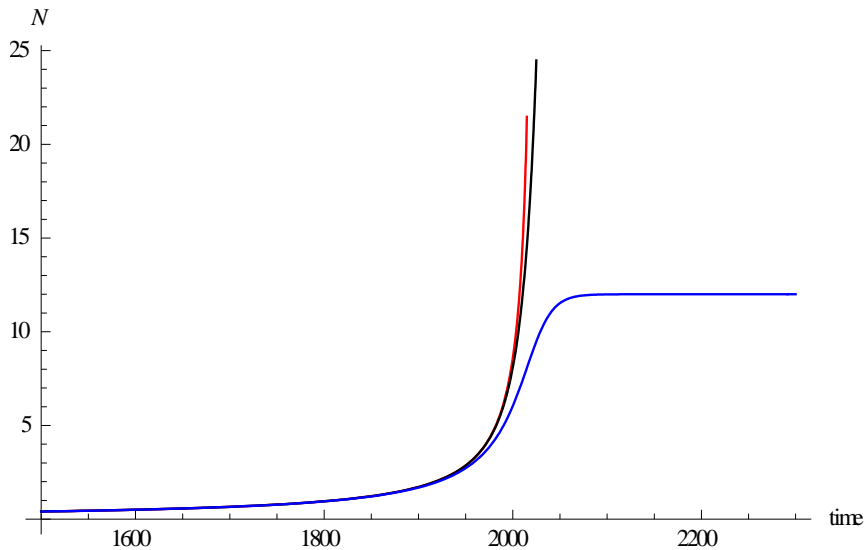

Figure 1. World population size, $N$ (in bln), dependently on time $t$: 1 (red) - models (3.1) with exponentially distributed rate; 2 (black) - model (3.4); 3 (blue) - inhomogeneous logistic model (3.8) with $C=12$ and $k=2$.

The subsequent transition from the Malthusian model to the inhomogeneous logistic model (see below) shows a transition from prolonged hyperbolical growth (the phase of "hyper-exponential" development) to the brief transitional phase of "nearly exponential" growth accomplished by a sharp increase of the variance of the reproduction rate to a stabilized regime, see Fig.2. It means that according to the last model humankind stay inhomogeneous forever; see discussion in s. 3.4.

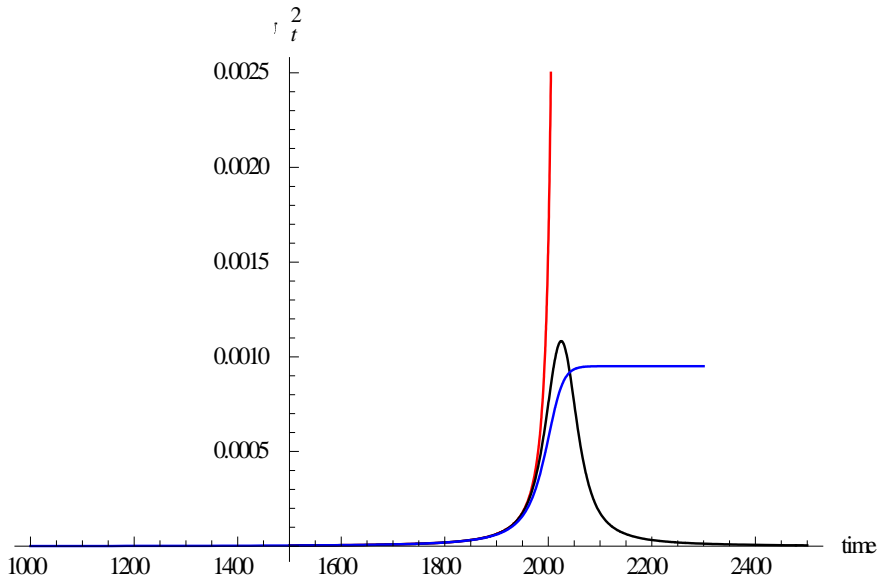

Figure 2. Variances $\sigma_t^2$ of the rate *a* dependently on time *t* for the models 1-3 (see Legend for Fig.1).

We conclude that the hyperbolic growth of the humankind for a long time period *was not an exclusive phenomenon* but obeyed the same laws as any heterogeneous biological population.

3.3. *Models of tree stand self-thinning*

The problem of dynamics of the tree number in a forest is one of the oldest and most important problems in forest ecology. A number of tree interactions, variations in genetic structure, and various environmental conditions affect the growth and death of trees in complex ways. It seems to be impossible to take into account all factors that impact death rates of trees in explicit form within the frameworks of a unit model.

A promising way to overcome these difficulties is to construct tree population models with distributed values of the mortality rate *a*. It was shown in [10], [12] that different formulas of forest stand self-thinning can be considered as solutions of inhomogeneous Malthus extinction model

$$\frac{dl(t,a)}{dt} = -acl(t,a), N(t) = \int_A l(t,a)da. \tag{3.5}$$

where $c$ is a scaling constant. Let us consider, for example, a known formula suggested by Hilmi for tree number $N(t)$ of even-age tree stand,

$$N(t) = N(0)\exp(a_0(e^{-ct} - 1)).$$

One can notice that this formula coincides with a well-known Gompertz function. It was shown in [10] that the last formula is an exact solution of (3.5) if the mortality rate $a$ has the Poisson distribution with average $a_0$ at the initial time point. In this case, the distribution of $a$ at any moment of time is also Poisson with the mean $E_t[a] = a_0 e^{-ct}$. Indeed, the mgf of the Poisson distribution $M(\lambda) = \exp(a_0(e^{-ct} - 1))$, and all assertions follow from formulas of s.2.

Thus, the Hilmi formula describes the Malthusian process of decreasing of the inhomogeneous population size; the population is divided into countable number of disjoint groups such that individuals from $i$-th group, $i = 0,1,2, ...$ have the mortality rate $ic$ and the initial size of the group is $l_0(i) = N(0)\exp(-a_0)a_0^i/i!$.

It is difficult to give an acceptable interpretation of this model with infinite number of groups. Individuals from groups with large $i$ have unrealistic large extinction rate and will be eliminated rapidly from the population; more of that, the initial size of such a group may be less than 1 with large $i$. Thus, it is reasonable to consider the subdivision of a tree population only to finite number of groups that corresponds to the truncated Poisson distribution.

Let parameter $a$ can assume only finite number of values $i = 0,1, ... k$. Then for all $t$ the distribution of the parameter $a$ is again the $k$-truncated Poisson distribution and the population size is

$$N(t) = N(0)c(k)\sum_{i=0}^{k}(\exp(-a_0)a_0)^i/i!. \tag{3.6}$$

This formula is biologically more meaningful then initial Hilmi formula with $k = \infty$; it allows to fit real data more precisely. More of that, it is possible now to choose the unknown number of different $a$-groups $k$ by best fitting, for different $k$, to the observed time series of the population sizes. Thus we can estimate the number of groups of trees having different "surviving levels". For examples, computations for "normal" pine tree stands give the estimation of $k$=7-10. The accuracy of data fitness is better then 5% of mean-square deviation $S$. Figure 3 shows the model solution and real data for normal pine planting of 1[st] class of quality ([26], Tables 130).

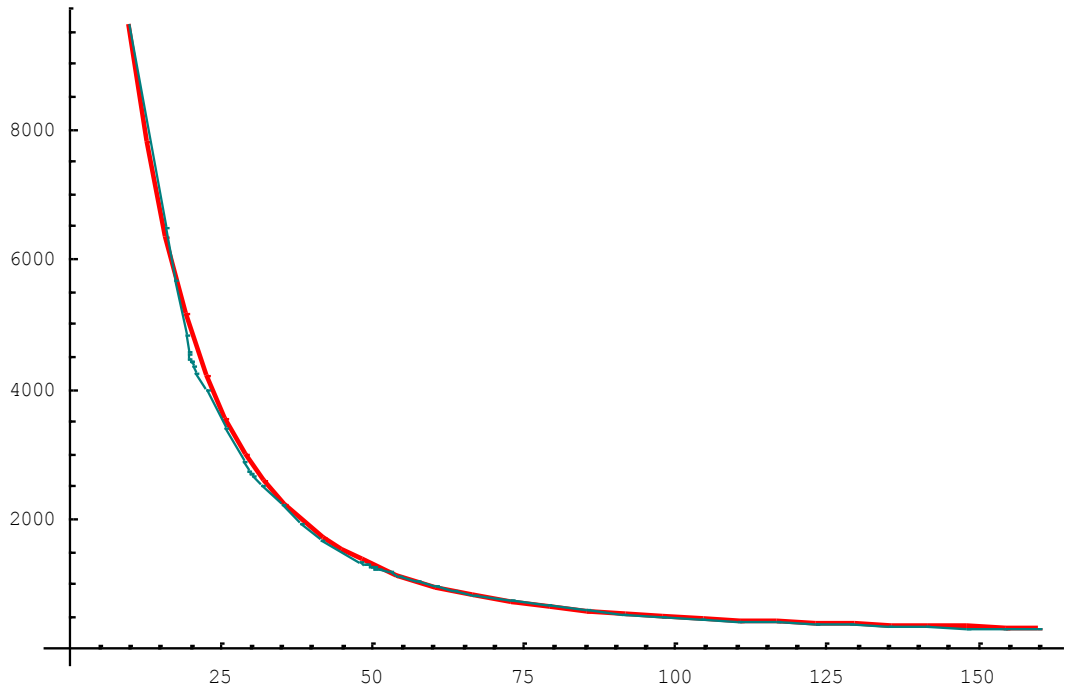

Figure 3. Data for pine planting of 1st class of quality (black) and modified Hilmi formula (3.1) (red) with *k*=9, $\alpha_0$ =3.57, *c*=0.02, *S*=4.2%.

Some other self-thinning formulas (Kayanus, Voropanov, the power formula) can be deduced and then modified by the same approach [10], [12].

3.4. ***Struggle for Existence and inhomogeneous logistic equation***

All populations have the capacity to grow exponentially under ideal conditions, and no population can grow exponentially forever – there are limits to growth. This generates the Malthusian Struggle for Existence [5]. A common opinion is that the struggle for existence results in the survival of the fittest. The following model shows that the situation is more complex.

A simple conceptual model for Malthusian Struggle for Existence is given by inhomogeneous logistic equation, which accounts both for free exponential growth and for recourse limitations:

$$\frac{dl(t;a,C)}{dt} = al(t;b,C)\left(1 - \frac{N}{C}\right) \qquad (3.7)$$

where $a$ is the Malthusian reproduction rate, which is assumed to be distributed, $C$ is the common caring capacity, and $N$ is the total population size. Notice that a more general logistic-type inhomogeneous model of the form

$$\frac{dl(t;a,C)}{dt} = al(t;b,C)(1-\frac{N}{C})^k, \text{ where } k > 0 \text{ is a constant} \tag{3.8}$$

is also of use.

Define the auxiliary variable by the equation

$$\frac{dq}{dt} = (1-\frac{N}{C})^k, q(0) = 0. \tag{3.9}$$

Then

$$l(t,a) = l(0,a)\exp(aq(t)); \tag{3.10}$$

the total size of the population is given by the formula

$$N(t) = N(0)M(q(t)) \tag{3.11}$$

and solves the logistic-like equation

$$dN/dt = E_t[a]N(1-\frac{N}{C})^k$$

where $M(\lambda)$ is the mgf of the initial distribution of the parameter $b$.

Hence, we have reduced inhomogeneous infinitely-dimensional logistic equation (3.7) to a single equation for $q(t)$ (3.9), (3.11). The keystone variable $q(t)$ can be considered as "internal time" such that the dynamics of inhomogeneous logistic model (3.7) with respect to the internal time (see (3.10)) is identical to the dynamics of the inhomogeneous Malthusian model with respect to the regular "external" time. The main difference is that in general $q(t)$ tends to $q^* < \infty$ as $t \to \infty$, which is the solution to the equation $M(q^*) = C/N(0)$.

Hence, the limit state of the inhomogeneous logistic model (3.7) coincides with the current state of the inhomogeneous Malthus model at the instant $q^*$. The limit stable population size and the limit distribution of the parameter $a$ are

$$N^* = N(0)M(q^*),$$

$$P^*(a) = P_0(a)\exp(q^*a)/M_0(q^*).$$

Let us emphasize a notable property of the inhomogeneous logistic model (3.7) with a distributed Malthusian parameter: it remains inhomogeneous at any instant and has a non-trivial limit distribution of the parameter at $t \to \infty$. Every clone that was present initially will be present in the limit stable state. Therefore, inhomogeneous logistic model illustrates the phenomenon of "survival of everybody" in the population, in contrast to Darwinian "survival of the fittest".

### 3.5. *Inhomogeneous Ricker equation*

Let us now demonstrate how to solve the inhomogeneous version of the well-known Ricker equation

$$\frac{dl(t;\beta,\mu)}{dt} = l(t;\beta,\mu)(\beta e^{-cN(t)} - \mu) \tag{3.12}$$

where $c$ is a constant, $\beta$ and $\mu$ are distributed parameters.

Let $M(\lambda_1, \lambda_2)$ be the mgf of the joint initial distribution of $\beta$ and $\mu$. Then the escort system reads

$$\frac{dq_1}{dt} = e^{-cN(0)}M(q_1,-t), q_1(0) = 0; \tag{3.13}$$

$$\frac{dq_2}{dt} = -1, q_2(0) = 0, \text{ hence } q_2(t) = -t.$$

Equation (3.13) can be solved at known mgf $M(\lambda_1, \lambda_2)$ and then the solution to (3.12) is given by $l(t;\beta,\mu) = l(0;\beta,\mu) \exp(\beta q_1(t) - \mu t)$; the total population size and the system distribution are given by the formulas

$$N(t) = N(0)M(q_1(t),-t),$$

$$P(t;\beta,\mu) = P(0;\beta,\mu) \exp(\beta q_1(t) - \mu t)/M(q_1(t),-t).$$

For example, let the parameters $\beta$ and $\mu$ be independent and exponentially distributed in $[0,\infty)$ with the means $s_1$ and $s_2$ at the initial instant. Then $M(q,-t) = \frac{s_1 s_2}{(s_1-q)(s_2+t)}$, and equation (3.13) for the auxiliary variable reads $\frac{dq_1}{dt} = \exp(-cN(0) \frac{s_1 s_2}{(s_1-q_1)(s_2+t)})$.

This equation has a stable state at $q_1 = s_1$. As $t \to \infty$, $q_1(t) \to s_1$, the total population size tends to infinity and the population density concentrates at the value $\mu = 0$ of the parameter $\mu$ and vanishes in any finite interval of values of the parameter $\beta$.

### 3.6. *The Fisher-Haldane-Wright equation*

One of the first replicator equations was introduced by R. Fisher in 1930 [3] for genotype evolution:

$$\frac{dp_a}{dt} = p_a(W_a - W), a = 1,\ldots n \tag{3.14}$$

where $W_a = \sum_b W_{ab}p_b$, $W = \sum_{a,b} p_a W_{ab} p_b$. Here $p_a$ is the frequency of the gamete $a$, $W_{ab}$ is the absolute fitness of the zygote $ab$. In mathematical genetics this equation is known as the Fisher-Haldane-Wright equation (FHWe) and sometimes is referred to as the main equation of mathematical genetics [25].

The matrix $\mathbf{W} = \{W_{ab}\}$ is symmetric and hence has the spectral representation

$$W_{ab} = \sum_{k=1}^{m} w_k h_k(a) h_k(b) \qquad (3.15)$$

where $w_k$ are non-zero eigenvalues and $h_k$ are corresponding orthonormal eigenvectors of the matrix $\mathbf{W}$; $m$ is the rank of $\mathbf{W}$. Then

$$W_a(t) = \sum_b W_{ab} p_b(t) = \sum_{k=1}^{m} \sum_{b=1}^{n} p_b(t) w_k h_k(a) h_k(b) = \sum_{k=1}^{m} w_k h_k(a) E_t[h_k].$$

The FHW-equation now reads

$$\frac{dp_a}{dt} = p_a \left( \sum_{k=1}^{m} w_k (h_k(a) E_t[h_k] - (E_t[h_k])^2) \right), a = 1, \ldots n. \qquad (3.16)$$

Consider the associated selection system (compare with (1.1) and (1.3)):

$$\frac{dl(t,a)}{dt} = l(t,a) \sum_{k=1}^{m} w_k h_k(a) E_t[h_k]. \qquad (3.17)$$

The range of values of the parameter $a$ is now a finite set, unlike in the previous examples. Define the function

$$M(\boldsymbol{\lambda}) = \sum_{a=1}^{n} \exp(\sum_{k=1}^{m} \lambda_k h_k(a) P(0,a)) = E_0[\exp(\sum_{k=1}^{m} \lambda_k h_k(a))].$$

Write out and solve the escort system of ODE

$$\frac{dq_i}{dt} = w_i E_0[h_i(a) \exp(\sum_{k=1}^{m} q_k h_k(a))] / E_0[\exp(\sum_{k=1}^{m} q_k h_k(a))], i = 1, \ldots m$$

These equations can be written in a more compact form

$$\frac{dq_i}{dt} = w_i \partial \ln M(\mathbf{q}) / \partial q_i. \qquad (3.18)$$

Then the solution to the selection system (3.17) is given by the formula

$l(t,a) = l(0,a) K(t,a)$ where $K(t,a) = \exp(\sum_{k=1}^{m} q_k(t) h_k(a))$;

the population size $N(t) = N(0) M(\mathbf{q}(t))$,

the values of regulators at $t$ moment $E_t[h_k] = \partial \ln M(\mathbf{q}) / \partial q_i$

and the current system distribution

$$P(t,a) = P(0,a) K(t,a) / E_0[K(t,a)] \qquad (3.19)$$

with $E_0[K(t,a)] = M(\mathbf{q}(t))$.

Formula (3.19) gives the solution to FHW-equation (3.14).

Technically the described approach is useful only if the rank of the fitness matrix $W$ is significantly smaller than its dimension, $m < n$. The approach is especially useful for infinitely dimensional system (3.14). Let us remark that, in general, the fitness matrix cannot be known exactly, but its elements can be well approximated by expression (3.15) with small $m$. For example, if $W_{ab} = w_a w_b$ for all $a, b$, then the initial multi-dimensional (or even infinitely-dimensional) system (3.14) is reduced to a single ODE. This case corresponds to a well-known example of a population in the Hardy-Weinberg equilibrium. See [18] for more general results.

## 4. Discrete time selection systems

### 4.1. Reduction theorem for maps

A similar theory can be developed for selection systems and replicator equations with discrete time [15]. Once again, let $l(t,\mathbf{a})$ be the population density at moment $t$; the discrete-time version of the model (1.1) of selection systems reads

$$l(t+1,\mathbf{a}) = l(t,\mathbf{a})w_t(\mathbf{a}), \tag{4.1}$$

$$w_t(\mathbf{a}) = \exp(\sum_{i=1}^{n} u_i(t,\mathbf{G}(t))\varphi_i(\mathbf{a})),$$

where again $\mathbf{G}(t) = G_1(t),...G_m(t)$, $G_i(t) = N(t)E_t[g_i]$, $N(t)$ is the total population size and $u_i(t,\mathbf{G})$ are continuous functions.

Define the current pdf $P_t(\mathbf{a}) = l(t,\mathbf{a})/N(t)$; it solves the discrete-time replicator equation

$$P(t+1,\mathbf{a}) = P(t,\mathbf{a})w_t(\mathbf{a})/E_t[w_t]. \tag{4.2}$$

Let us formulate the reduction theorem for selection system (4.1) (which was proven in [15] for less general model). Denote

$$U_i(t) = \sum_{j=0}^{t} u_i(j,\mathbf{G}(j)), \text{ and}$$

$$W(t,\mathbf{a}) = \prod_{k=0}^{t} w_k(\mathbf{a}) = \exp(\sum_{i=1}^{n} U_i(t)\varphi_i(\mathbf{a})). \tag{4.3}$$

Consider the vector-function $\mathbf{U}(t) = (U_1(t),...U_n(t))$. Suppose that the initial distribution $P(0,\mathbf{a})$ is given and $\Phi(r;\lambda)$ is the generating functional (2.1).

**Theorem 4.** The total population size, the regulators and the system distribution can be computed recurrently with the help of the following system:

$$N(t) = N(0)\Phi(1; \mathbf{U}(t-1));$$

$$G_i(t) = N_0\Phi(g_i; \mathbf{U}(t-1));$$

$$P(t,\mathbf{a}) = P(0,\mathbf{a})W(t-1,\mathbf{a})/E_0[W(t-1,\mathbf{a})] \tag{4.4}$$

Theorem 4 reduces model (4.1)-(4.2) (which in general is infinitely dimensional) to a finite system of recursive equations. Formulas (4.3)-(4.4) give the solution to replicator equation (4.2).

*4.2. Inhomogeneous logistic map*

A well-known logistic map is of the form $N_{t+1} = \lambda N_t(1-N_t)$, $0 < \lambda < 4$ and $0 \leq N_t \leq 1$. The following inhomogeneous version of this map was solved in [15], example 7.2:

$$l(t+1,a) = \lambda a l(t,a)(1-N_t)$$

where $a$ is the distributed parameter, $0 < a < 1$. Denote $w_t(a) = \lambda a(1-N_t)$.

Let $W(t,a) = \prod_{k=0}^{t} w_k(a) = \lambda^{t+1}\exp((t+1)\ln a + U(t))$,

where $U(t) = \sum_{k=1}^{t}\ln(1-N_k)$. Then, according to (4.4) the current pdf

$$P(t,a) = P(0,a)W(t-1,a)/E_0[W(t-1,a)] = P(0,a)a^t/E_0[a^t]. \tag{4.5}$$

The inhomogeneous logistic model is reduced to the system

$$N_{t+1} = E_t[w]N_t = \lambda E_t[a]N_t(1-N_t), \tag{4.6}$$

$$E_t[a] = E_0[a^{t+1}]/E_0[a^t].$$

The current distribution (4.5) of the model has a simple form. In contrast, the behavior of trajectories may be very complex. For example, let $P_0(a)$ be the Beta-distribution in [0,1] with parameters $\alpha, \beta$. Then $E_t[a] = (t+\alpha)/(t+\alpha+\beta)$. Choosing an appropriate value of $0 < \lambda < 4$, we can observe (as $t \to \infty$) any possible behavior of the plain logistic model as the final dynamical behaviors of the inhomogeneous logistic model. In particular, at $\lambda = 4$ almost all cycles of Feigenbaum's cascade appear in the course of time and realize as parts of a single trajectory, as a result of the "inner" bifurcations of the inhomogeneous logistic model. The trajectory $\{N_t\}_0^\infty$ mimics the bifurcation diagram of the plain logistic map. Figure 4 illustrates this assertion.

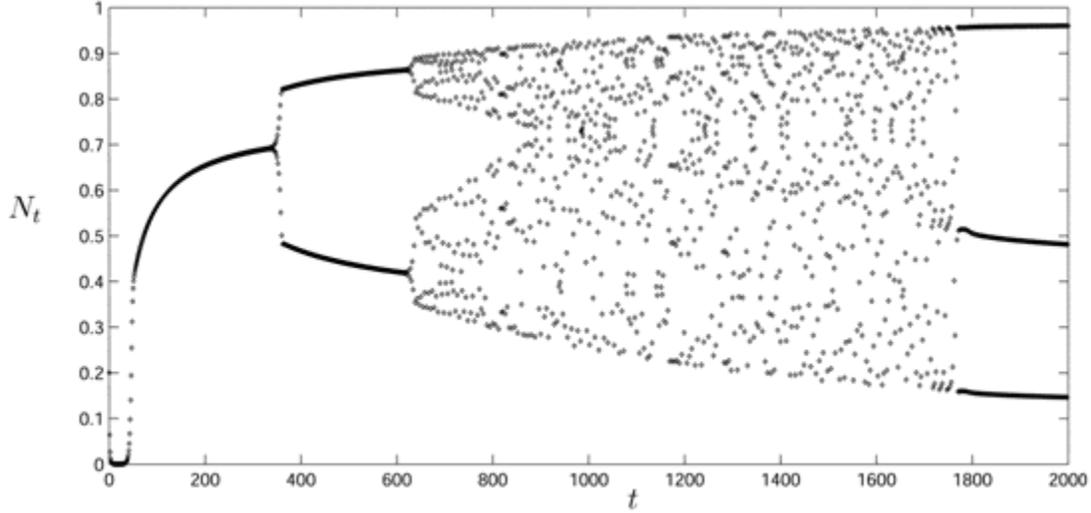

Figure.4. The trajectory of total population size for inhomogeneous logistic model with Beta-distributed parameter $a$ ($\lambda = 4, E_0[a] = 0.1, Var_0[a] = 0.005$).

*4.3. Inhomogeneous Ricker' map*

The Ricker' model takes into account the population size regulation of the reproduction rate using a more appropriate way then does the logistic map. Consider the inhomogeneous version of the Ricker' model, which depends on two distributed parameters $a$ and $b$ (see [15], example 7.3):

$$l(t+1;a,b) = l(t;a,b)\exp(a - bN_t). \qquad (4.7)$$

Denote for brevity $U(t) = \sum_{k=0}^{t} N_k$, $w_t = \exp(a - bN_t)$. Then $l(t+1;a,b) = l(0;a,b)W(t;a,b)$

where $W(t;a,b) \equiv \prod_{k=0}^{t} w_k = \exp((t+1)a - bU(t))$.

Let $M(\lambda_1, \lambda_2) = \int_A \exp(\lambda_1 a + \lambda_2 b)P_0(a,b)dadb$ be the moment generating function of the joint initial distribution of the parameters $a, b$. Recall that $\Phi(1;\lambda_1,\lambda_2) = M(\lambda_1,\lambda_2)$ and

$E_0[W(t;a,b)] = M(t+1,-U(t))$.

Applying Theorem 4 we obtain

$N_t = N_0 M(t, -U(t-1))$,

$P(t; a, b) = P(0; a, b) \exp(at - bU(t-1)) / M(t, -U(t-1))$.  (4.8)

The function $U(t)$ can be computed by the recurrent equation

$U(t+1) = U(t) + N_0 M(t+1, -U(t))$.

These formulas completely solve the inhomogeneous Ricker' model. In order to illustrate the peculiarities of the model dynamics, let us suppose that the parameter $b > 0$ is fixed and the initial distribution of $a$ is the $\Gamma$-distribution with the parameters $(s, k)$. Then (see [15].[16]) $P_t(a)$ is again the $\Gamma$-distribution with the parameters $(s, k+t)$,

$P(t; a) = s^{k+t} a^{k+t-1} \exp(-sa) / \Gamma(k+t) = (sa)^t \Gamma(k) / \Gamma(k+t) P(0; a)$

and $N_{t+1} = N_t \exp(-bN_t)(k+t)/s$.

These formulas show that the coefficient $(k+t)/s$, which determines the dynamics of the Ricker' model, increases indefinitely with time. As a result, according to the theory of the plain homogeneous Ricker' model, after a period of monotonic increase, cycles of period 2, then of period 4, and then almost all cycles of Feigenbaum's cascade appear and realize as parts of a single trajectory, see [15] for details. The trajectories $\{N_t\}_0^\infty$ mimics the bifurcation diagram of the plain Ricker' model, see Fig.5.

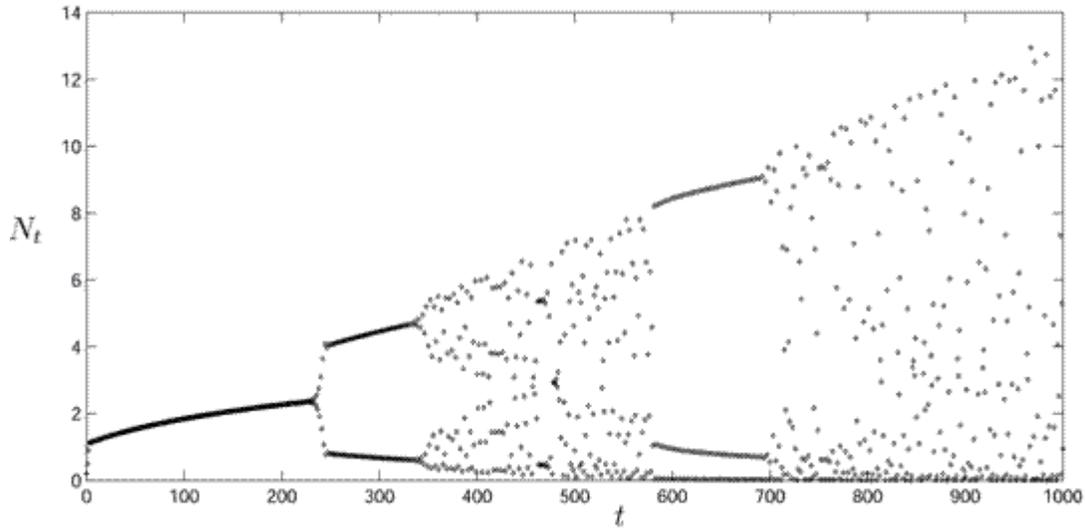

Figure 5. The trajectory of total population size for inhomogeneous Ricker' model with $\Gamma$ − distributed parameters $a$ ($\lambda_0 = 1, E_0[a] = 3, Var_0[a] = 0.1$).

### 4.4. Selection in Natural Rotifer Community

The mathematical model of zooplankton populations under toxicant exposure was suggested in [24] and studied systematically in [2]. The model depends on the parameters *a,* characterizing the environment quality, and γ, which is the species-specific parameter. The model was given by the equation

$$N_{t+1} = N_t \exp(-a + \frac{1}{N_t} - \frac{\gamma}{N_t^2}).$$

Let us consider the model of a community that consists of different rotifer populations (see [14]); individuals inside the populations may have different reproduction capacities under constant toxicant exposure. The model is of the form (4.1), $l(t+1,\mathbf{a}) = l(t,\mathbf{a})w_t(\mathbf{a})$, where $\mathbf{a} = (a,\gamma)$ and

$$w_t(a,\gamma) = \exp(-a + \frac{1}{N_t} - \frac{\gamma}{N_t^2}).$$

Then

$$W(t;a,\gamma) = \exp(-(t+1)a + \sum_{k=0}^{t} 1/N_k - \gamma \sum_{k=0}^{t} 1/N_k^2).$$

Let $M(\lambda_1, \lambda_2)$ be the mgf of the initial distribution of $a$ and $\gamma$. Then

$$E_0[W(t;a,\gamma)] = \exp(\sum_{k=0}^{t} 1/N_k) M(-(t+1), -\sum_{k=0}^{t} 1/N_k^2),$$

$$N_t = N_0 \exp(\sum_{k=0}^{t-1} 1/N_k) M(-t, -\sum_{k=0}^{t-1} 1/N_k^2);$$

$$P(t;a,\gamma) = P(0;a,\gamma)\exp(-ta - \gamma\sum_{k=0}^{t-1}1/N_k^2)/M(-t,-\sum_{k=0}^{t-1}1/N_k^2).$$

These formulas completely solve the inhomogeneous model of rotifer community.

Notice that inhomogeneous logistic and Ricker' maps show transitions from deterministic to chaotic behaviors; in contrast, the model of rotifer community shows the transition from almost chaotic to deterministic behavior. The trajectory $N_t$ may have a very complex transition regime from the initial to the final behavior; see Figure 6.

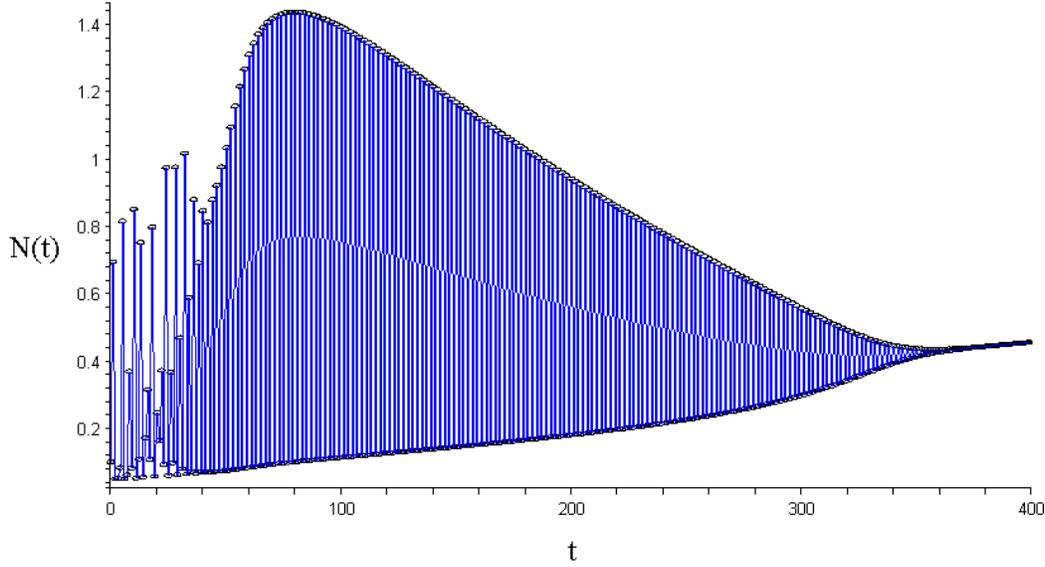

Figure 6. The trajectory of total population size for inhomogeneous rotifer' model where independent parameters $a, \gamma$ are both $\Gamma-$ distributed; $E_0[a] = 4, Var_0[a] = 0.01;\ E_0[\gamma] = 0.06, Var_0[\gamma] = 0.03$.

## 5. Inhomogeneous models of communities

### 5.1. Reduction Theorem

Consider the model of a community consisting from $r$ interacting populations. We suppose again that every individual is characterized by their own value of vector-parameter $\mathbf{a}$. Let $l^j(t,\mathbf{a})$ be the density of $j$-th population at moment $t$. In this section we consider the model of an inhomogeneous community where the reproduction rates can depend on current characteristics of every population in the community composing a "regulator". Formally, we consider the set of $m$ regulators, each of which is the $r$-dimension vector-function $G_i(t) = (G_i^1(t),...G_i^r(t))$, $i = 1,...n$ where:

$$G_i^{\,j}(t) = \int_A g_i(\mathbf{a}) l^{\,j}(t,\mathbf{a}) d\mathbf{a} \tag{5.1}$$

Each regulator corresponds to appropriate weight function $g_i$. A finite set of the regulators corresponds to each specific model; we denote this set as $\mathbf{G}(t) = (G_1(t),...G_n(t))$.

The current population sizes $N^j(t) = \int_A l^{\,j}(t,\mathbf{a}) d\mathbf{a}$ compose the regulator of a special importance, $N(t) = (N^1(t),...N^r(t))$. We assume that $N(t)$ is included in the set of the model' regulators. The distribution of $j$-th population in the community is by definition $P^j(t,\mathbf{a}) = l^{\,j}(t,\mathbf{a}) / N^j(t)$.

The model of inhomogeneous community considered here (see [17]) is of the form:

$$dl^{\,j}(t,\mathbf{a})/dt = l^{\,j}(t,\mathbf{a}) F^{\,j}(t,\mathbf{a}) \tag{5.2}$$

$$F^{\,j}(t,\mathbf{a}) = \sum_{i=1}^{n} u_i^{\,j}(t,\mathbf{G}) \varphi_i(\mathbf{a}) \tag{5.3}$$

where the functions $u_i^{\,j}$ can be specific for each trait and each population. The initial pdf-s $P^j(0,\mathbf{a})$ and the initial population sizes $N^j(0)$ are assumed to be given. The current pdf $P^j(t,\mathbf{a})$ solves the replicator equation:

$$dP^j(t,\mathbf{a})/dt = P^j(t,\mathbf{a})(F^{\,j}(t,\mathbf{a}) - E_t^{\,j}[F^{\,j}]) \tag{5.4}$$

The theory for inhomogeneous community model (5.1)-(5.4) is similar to the theory presented in Section 3 for inhomogeneous populations up to more complex technical details. Theorem 5 reduces complex model (5.1)-(5.3) to an escort system of ordinary non-autonomic equations of dimension $r \times n$ and gives the solution to replicator equation (5.4). Introduce the keystone variables as the solution to the following Cauchy problem:

$$dq_i^{\,j}/dt = u_i^{\,j}(t,\mathbf{G}_i^*(t)), \qquad q_i^{\,j}(0) = 0, \qquad j = 1,...r, \qquad i = 1,...n. \tag{5.5}$$

Here:

$$\mathbf{q}^{\,j}(t) = (q_1^{\,j}(t),...q_n^{\,j}(t)), \quad G_i^{*j}(t) = N^j(0)\Phi^{\,j}(g_i; \mathbf{q}^{\,j}(t)),$$

where $\Phi^{\,j}(r;\lambda)$ is the generating functional (2.5) for the initial distribution $P^j(0,\mathbf{a})$; in particular, $N^{*j}(t) = N^j(0)\Phi^{\,j}(1;\mathbf{q}^{\,j}(t))$.

Denote $K_t^{\,j}(\mathbf{a}) = \exp(\sum_{i=1}^{n} q_i^{\,j}(t) \varphi_i(\mathbf{a}))$.

**Theorem 5 [17]**. *Suppose that Cauchy problem (5.5) has a unique global solution at $t \in [0,T)$, $0 < T \leq \infty$. Then the functions:*

$$l^j(t,\mathbf{a}) = l^j(0,\mathbf{a})K_t^j(\mathbf{a})$$

$$G_i^j(t) = N^j(0)\Phi^j(g_i;\mathbf{q}^j(t))$$

$$N^j(t) = N^j(0)\Phi^j(1;\mathbf{q}^j(t))$$

*satisfy system (5.1)- (5.3) at $t \in [0,T)$. The pdf:*

$$P^j(t,\mathbf{a}) = P^j(0,\mathbf{a})K_t^j(\mathbf{a})/E_0[K_t^j] = P^j(0,\mathbf{a})K_t^j(\mathbf{a})/\Phi^j(1;\mathbf{q}^j(t)) \tag{5.6}$$

*solves the replicator equations (5.4).*

Let us apply the general theory to some classical models of biological communities consisting of interacting inhomogeneous populations.

*5.2. Inhomogeneous prey-predator Volterra' model*

The classical prey-predator Volterra' model in its simplest form reads

$$dx/dt = a_1 x - a_2 xy, \tag{5.7}$$

$$dy/dt = -a_3 y + a_4 xy$$

where $x(t)$ and $y(t)$ denote prey and predator densities, $a_1$ is the reproduction rate of the prey population, $a_2$ is the per capita rate of the consumption of prey by the predators, $a_3$ is the death rate of the predator, and $a_4/a_2$ is the fraction of prey biomass, which is converted into predator biomass.

Let us consider the inhomogeneous version of this classical model supposing that parameters $a_1$, $a_2$, and $a_3$ are distributed and the ratio $a_4/a_2$ is fixed (and hence could be chosen equal to 1). We also suppose that the reproduction and death processes are specific for each subpopulation, while the consumption is driven by the interaction of the prey (predator) subpopulation with the entire predator (prey) population. Let $x(t;a_1,a_2)$, $y(t;a_3)$ be the densities of the prey and predator populations over parameters $a_1, a_2$ and $a_3$ correspondingly, and $X(t) = \int_A x(t;a_1,a_2)da_1 da_2$, $Y(t) = \int_A y(t;a_3)da_3$ be the total sizes of the populations. The initial population sizes and initial distributions $P^1(0;a_1,a_2), P^2(0;a_2)$ are assumed to be given. The total rate of consumption is equal to

$Y(t)\int_A a_2 x(t;a_1,a_2)da_1 da_2 = Y(t)X(t)E^1{}_t[a_2]$. Assuming the "proportional distribution" of prey among the predators we can write the inhomogeneous version of Volterra' model in the form

$$dx(t;a_1,a_2)/dt = x(t;a_1,a_2)(a_1 - a_2 Y(t)), \qquad (5.8)$$

$$dy(t;a_3)/dt = y(t;a_3)(G(t) - a_3)$$

where $G(t) = \int_A a_2 x(t;a_1,a_2)da_1 da_2 = X(t)E_t^1[a_2]$.

Theorem 5 gives a method to study this model; the principal step is a reduction of the complex system (5.1)-(5.3) to the escort system of ODE for keystone variables. It is instructive to deduce the escort system and the main results informally to clarify the main idea of the method in application to community models.

It is natural to suppose that the parameter $a_3$ is stochastically independent on the parameters $a_1, a_2$. Let $M^1(\lambda_1, \lambda_2)$ be the mgf of the initial joint distribution of the parameters $a_1, a_2$, and $M^2(\lambda_3)$ be the mgf of the initial distribution of the parameter $a_3$.

Introduce the auxiliary keystone variables $q_1(t)$, $q_2(t)$ as a solution to the Cauchy problem

$$dq_1/dt = Y(t),$$

$$dq_2/dt = G(t) = X(t)E_t^1[a_2],$$

$$q_1(0) = q_2(0) = 0.$$

Then system (5.8) can be written formally as

$$dx(t;a_1,a_2)/dt = x(t;a_1,a_2)(a_1 - a_2 dq_1/dt),$$

$$dy(t;a_3)/dt = y(t;a_3)(dq_2/dt - a_3).$$

Its solution is

$$x(t;a_1,a_2) = x(0;a_1,a_2)\exp(a_1 t - a_2 q_1(t)),$$

$$y(t;a_3) = y(0;a_3)\exp((q_2(t) - a_3 t)).$$

Now we can express all values of interest with the help of the mgf-s of the initial distributions and the auxiliary variables:

$$X(t) = X(0)\int_A \exp(a_1 t - a_2 q_1(t))P^1(0;a_1,a_2)da_1 da_2 = M^1(t,-q_1(t))$$

$$Y(t) = Y(0)\int_A \exp((q_2(t) - a_3 t))P^2(0;a_3)da_3 = \exp(q_2(t))M^2(-t) \qquad (5.9)$$

$$P^1(t;a_1,a_2) = x(t;a_1,a_2)/X(t) = \exp(a_1 t - a_2 q_1(t))/M^1(t,-q_1(t))P^1(0;a_1,a_2),$$

$$P^2(t;a_3) = y(t;a_3)/Y(t) = \exp(-a_3 t)/M^3(-t)P^2(0;a_3), \qquad (5.10)$$

$$E_t^1[a_2] = (\int_A a_2 \exp(a_1 t - a_2 q_1(t))P^1(0;a_1,a_2)da_1 da_2)/M^1(t,-q_1(t)), \text{ hence}$$

$$E_t^1[a_2]X(t) = \frac{\partial}{\partial \lambda_2} M^1(\lambda_1,\lambda_2)\Big|_{\lambda_1=t,\lambda_2=-q_1}.$$

Finally, we obtain a closed system of non-autonomous equations

$$dq_1/dt = \exp(q_2(t))M^2(-t),$$

$$dq_2/dt = \frac{\partial}{\partial \lambda_2} M^1(t_1,-q_1).$$

Now that we have a solution to the Cauchy problem for this system with zero initial values, we can get explicit formulas for total populations' sizes (5.9) and current distribution of the parameters (5.10), which completely solve the problem. In particular, the current mean values of the parameters

$$E^1{}_t[a_1] = \frac{\partial}{\partial \lambda_1} \ln(M^1(t,-q_1(t))), \qquad (5.11)$$

$$E^1{}_t[a_2] = \frac{\partial}{\partial \lambda_2} \ln(M^1(t,-q_1(t))),$$

$$E^2{}_t[a_3] = \frac{\partial}{\partial \lambda_3} \ln(M^2(-t)).$$

One can check that the obtained formulas coincide with the formulas, which follow from Theorem 5.

Integrating the equations of system (5.8) over the parameters we obtain the system

$$dX/dt = X(E^1{}_t[a_1] - E^1{}_t[a_2]Y), \qquad (5.12)$$

$$dY/dt = Y(E_t^1[a_2]X - E_t^2[a_3]).$$

These equations for total sizes of inhomogeneous populations have the same form as the initial Volterra' system (5.7); the difference is that now the parameter values are not constants but vary over time according to formulas (5.11). The phase-parametric portrait of "homogeneous" Volterra' model is well known (see, e.g., [1]). The dynamics of system (5.12) is determined by the parametric point $(E^1{}_t[a_1], E^1{}_t[a_2], E^2{}_t[a_3])$, which moves across the parametric portrait of model (5.7). This phenomenon, which may be referred to as "traveling across the parametric portrait of a homogeneous model" is a common feature of corresponding inhomogeneous models. We have observed it on the example of discrete-time models in s.5 ("traveling along the bifurcation diagram" of corresponding maps). For Volterra-type model of two inhomogeneous populations with logistic reproduction rates and ratio-dependent predator functional response the phenomenon was studied in detail in [13].

*8.3. Competition of two inhomogeneous populations*

The dynamics of two populations competing for a common resource can be described by the following logistic-like model (see, e.g., [1], ch.4):

$$dx/dt = ax(1-(x+\alpha y)/A),$$
$$dy/dt = by(1-(y+\beta x)/B)$$

where $A, B$ are the capacities of the ecological niches for both populations, and $\alpha, \beta$ are the coefficients of interspecies competitions. A more general Allee-like model has a form

$$dx/dt = ax((x-L)(A-x)-\alpha y), \qquad (5.13)$$
$$dy/dt = by((y-M)(B-y)-\beta x)$$

where $L, M$ are the lower threshold sizes for both populations.

Consider the inhomogeneous versions of these models, supposing that the reproduction rates $a, b$ are distributed, and the competition is defined by the total sizes $X(t), Y(t)$ of the population.

Then instead of the logistic model we obtain the following model

$$dx(t,a)/dt = ax(t,a)(1-(X(t)+\alpha Y(t))/A), \qquad (5.14)$$
$$dy(t,b)/dt = by(t,b)(1-(Y(t)+\beta X(t))/B)$$

and correspondingly

$$dx(t,a)/dt = ax(t,a)((X(t)-L)(A-X(t))-\alpha Y(t)),$$
$$dy(t,b)/dt = by(t,b)((Y(t)-M)(B-Y(t))-\beta X(t))$$

instead of the Allee-type model. Both systems have a form

$$dx(t,a)/dt = ax(t,a)(u(X(t))-\alpha Y(t)), \qquad (5.15)$$
$$dy(t,b)/dt = by(t,b)(v(Y(t))-\beta X(t))$$

where $u, v$ are appropriate functions.

Let $P^1(0;a)$ and $P^2(0;b)$ be the initial distributions of the Malthusian rates $a,b$ and $M^1(\lambda)$, $M^2(\lambda)$ be corresponding mgf-s. In order to study model (5.15), we apply Theorem 5 and consider the 4-dimension escort system:

$$dq_1^1/dt = u(X(0)M^1(q_1^1 - \alpha q_2^1)), \qquad (5.16)$$

$$dq_2^1/dt = Y(0)M^2(q_1^2 - \beta q_2^2),$$

$$dq_1^2/dt = v(Y(0)M^2(q_1^2 - \beta q_2^2)),$$

$$dq_2^2/dt = X(0)M^1(q_1^1 - \alpha q_2^1),$$

$$q_1^1(0) = q_2^1(0) = q_1^2(0) = q_2^2(0) = 0.$$

Suppose that Cauchy problem (5.16) has a unique global solution at $t \in [0,T)$, $0 < T \leq \infty$.

Define $K_t^1(a) = \exp(a(q_1^1(t) - \alpha q_2^1(t)))$, $K_t^2(b) = \exp(b(q_1^2(t) - \beta q_2^2(t)))$.

Then the solution to model (5.15) is

$$x(t,a) = x(0,a)K_t^1(a) = x(0,a)\exp(a(q_1^1(t) - \alpha q_2^1(t))),$$

$$y(t,b) = y(0,b)K_t^2(a) = y(0,b)\exp(b(q_1^2(t) - \beta q_2^2(t)))$$

$$X(t) = X(0)M^1(q_1^1(t) - \alpha q_2^1(t)),$$

$$Y(t) = Y(0)M^2(q_1^2(t) - \beta q_2^2(t)).$$

The mean values of the Malthusian rates at $t$ moment

$$E^1_t[a] = \frac{\partial}{\partial \lambda} \ln(M^1(q_1^1(t) - \alpha q_2^1(t))), \qquad (5.17)$$

$$E^2_t[b] = \frac{\partial}{\partial \lambda} \ln(M^2(q_1^2(t) - \beta q_2^2(t))).$$

The current pdf-s

$$P^1(t;a) = P^1(0;a)\exp(a(q_1^1(t) - \alpha q_2^1(t)))/M^1(q_1^1(t) - \alpha q_2^1(t)), \qquad (5.18)$$

$$P^2(t;b) = P^2(0;b)\exp(b(q_1^2(t) - \beta q_2^2(t)))/M^2(q_1^2(t) - \beta q_2^2(t)).$$

A simpler logistic model (5.14) can be reduced to a two-dimension escort system

$$dq^1/dt = X(0)M^1(t - (q^1 + \alpha q^2)/A),$$

$$dq^2/dt = Y(0)M^2(t - (\beta q^1 + q^2)/B),$$

$$q^1(0) = q^2(0) = 0.$$

The solution to model (5.14) is given by the following formulas:

$$x(t,a) = x(0,a)\exp(a(t - (q^1(t) + \alpha q^2(t)/A))),$$

$$y(t,b) = y(0,b)\exp(b(t - (\beta q^1(t) + q^2(t)/B))),$$

$$X(t) = X(0)M^1(t - (q^1(t) + \alpha q^2(t))/A),$$

$$Y(t) = Y(0)M^2(t - (\beta q^1(t) + q^2(t))/B).$$

The current pdf-s

$$P^1(t;a) = P^1(0;a)\exp(a(t - (q^1(t) + \alpha q^2(t))/A))/M^1(t - (q^1(t) + \alpha q^2(t))/A),$$

$$P^2(t;b) = P^2(0;b)\exp(b(t - (\beta q^1(t) + q^2(t))/B))/M^2(t - (\beta q^1(t) + q^2(t))/B)$$

The mean values of the Malthusian rates at moment $t$ are

$$E^1_t[a] = \frac{\partial}{\partial \lambda}\ln(M^1(t - (q^1(t) + \alpha q^2(t))/A)),$$

$$E^2_t[b] = \frac{\partial}{\partial \lambda}\ln(M^2(t - (\beta q^1(t) + q^2(t))/B)).$$

Other examples of inhomogeneous community models can be found in [13], [17], [22].

## 6. *Preventing the tragedy of the commons through punishment of over-consumers and encouragement of under-consumers*

The conditions that can lead to the exploitative depletion of a shared resource, i.e, the tragedy of the commons [6], can be reformulated as a game of prisoner's dilemma: while preserving the common resource is in the best interest of the group, over-consumption is in the interest of each particular individual at any given point in time. One way to try and prevent the tragedy of the commons is through infliction of punishment for over-consumption and/or encouraging under-consumption, thus selecting against over-consumers. The effectiveness of various punishment functions in an evolving system can be evaluated within a framework of a parametrically heterogeneous consumer-resource model.

Since reduction of the common 'carrying capacity' can indeed lead to population collapse, this model allows to model effectively in a conceptual framework the question of the effects of over-consumption on the survival of the population. Suppose that different individuals can subtract from the common recourse $z(t)$, which defines the dynamical carrying capacity of the population, at different rates $c$ or they can contribute to the common dynamical carrying capacity with the rate $1 - c$. Denote $x_c$ a set of consumers that are characterized by the same value of $c$. Each individual is punished or rewarded according to a function $f(c)$ that directly affects the fitness of each consumer-producer such that each clone is immediately punished for over-consumption if $c > 1$ or rewarded for under-consumption when $c < 1$. The final model (a version of the model of [21]) is of the form

$$\frac{dx_c}{dt} = rx_c(t)(c - \frac{N(t)}{kz(t)}) + x_c(t)f(c),$$
$$\frac{dz}{dt} = \gamma - \delta z(t) + e\frac{N(t)(1 - E^t[c])}{N(t) + z(t)}$$
(6.1)

Introduce a keystone variable $q(t)$ by the equation $\frac{dq}{dt} = \frac{N}{kz}$, so that one may rewrite the first equation in the following form:

$$\frac{dx_c}{dt} = rx_c(t)(c - \frac{dq}{dt}) + x_c(t)f(c).$$

Then $x_c(t) = x_c(0)\exp(-q(t) + t(rc + f(c)))$.

Total population size is then given by

$$N(t) = N(0)\exp(-q(t)) \int_c \exp(t(rc + f(c))P(0,c)dc,$$
(6.2)

the current mean value

$E^t[c] = \int_c cP(t,c)dc$ where

$$P(t,c) = \int_c c\exp(t(rc + f(c))P(0,c)dc / \int_c \exp(t(rc + f(c))P(0,c)dc.$$
(6.3)

We have reduced the initial complex model (6.1) to two-dimensional system

$$\frac{dz}{dt} = \gamma - \delta z + e\frac{N(1 - E^t[c])}{N + z}$$
$$\frac{dq}{dt} = \frac{N}{kz}$$
(6.4)

where $N$ and $E^t[c]$ are defined by formulas (6.2)-(6.3).

The model was studied analytically in [19] without incorporating punishment/reward for over/under consumption. As the value of $c$ increases, the population goes through a series of transitional regimes from sustainable coexistence with the resource to oscillatory regime to eventually committing evolutionary suicide through decreasing the common carrying capacity to a level that can no longer support the population.

The effectiveness of three types of punishment functions was evaluated in[20]: moderate punishment, $f(c) = a(1-c)/(1+c)$; severe punishment, $f(c) = a(1-c)^3$, where the parameter $a$ denotes the severity of implementation of punishment on individuals with the corresponding value of parameter $c$, and function of the type, $f(c) = r(1-c^b)$, which allows to separate the influence of reward for under-consumption, primarily accounted for with parameter $r$, and punishment for overconsumption, accounted for with parameter $b$. It was found that the effectiveness of punishment depend not only on intrinsic parameter values of the system but also on the initial composition of the population. When the punishment imposed is moderate, overconsumption could be avoided only when the value of $a$ was very high, i.e. when punishment is imposed very severely. The value of $a$ that was necessary for successful management of over-consumers varied depending on different initial distributions, indicating that in order to be able to prevent the tragedy of the commons, one needs to evaluate not only the type of punishment and the severity of its enforcement but also match it to the composition of the population, since one level of punishment can be effective for one distribution of clones within a population of consumers and not another.

The same set of numerical experiments was conducted for the severe punishment\generous reward function. Then, the value of $a$ that would correspond to successful restraint of over-consumers was much lower than in the previous case for all initial distributions considered here. The system was able to support individuals with higher values of parameter $c$ present in the initial population than in the previous case. In some cases brief periods of oscillatory transitional dynamical behavior before the system collapsed can be observed.

The punishment function of the type $f(c) = r(1-c^b)$ allows one to account the intensity of punishment and reward by parameters $b$ and $r$ respectively. In order to evaluate the expected effectiveness of the punishment/reward system one needs to not only adjust parameters $b$ and $r$ to each particular case considered but also be able to evaluate the expected range of parameter $c$, since one level of punishment may be appropriate for one set of initial conditions but not another. Moreover, the time to collapse varies depending on the initial distribution of the clones within the population, and the higher the frequency of over-consumers is in the initial population, the worse the prognosis.

Figure 7 shows the dynamics of the total population size and total resource with respect to different values of $a$ (different levels of severity of imposed punishment). One can see that successful management of overconsumers was possible only when punishment implementation was very high.

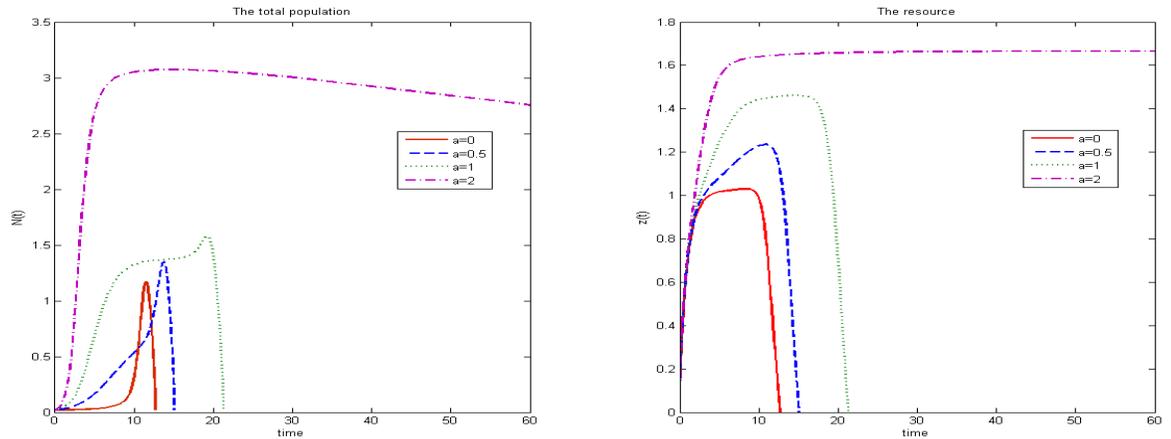

Fig.7. The model with severe punishment, $f(c) = a(1-c)^3$ and truncated exponential distribution of the parameter $c$. Dynamics of the total population size (left) and total resource (right) with respect to different values of $a$.

Overall, one can conclude that the tragedy of the commons can be prevented if the punishment of over-consumers increases *non-linearly* faster against the rate of over-consumption.

### Discussion

An overview of a general approach to a wide class of selection systems and replication equations is presented. The developed theory allows reducing complex inhomogeneous models to "escort systems" of ODEs for auxiliary "keystone" variables that, in many cases, can be investigated analytically. Notice that even if the analytical solution of the escort system is not available, numerically solving the Cauchy problem for the resulting system of ODEs is much simpler than studying the initial problem numerically. It allows computing (in many cases, explicitly) all statistical characteristics of interest of the initial selection system and visualizing evolutionary trajectories. The approach was developed both to continuous time and discrete time selection systems and replicator equations. The considered examples show how different the global dynamics of a selection system can be depending on the initial distribution even within the frameworks of the same dynamical model.

The developed method was systematically applied to different inhomogeneous models of populations and communities. Analytical solutions of the considered models can provide new biological insights beyond the computer simulations. The derived explicit solutions may be helpful and necessary in order to be able to completely study corresponding models, which belong to different areas of mathematical biology. Applications to some ecological problems [10], [14], [15], global demography [11], cancer modeling [13], epidemics in heterogeneous populations [18], ecological niche construction [19], modeling of the tragedy of commons [20], epidemiological modeling [22], [23], etc. were published recently. It is my hope that the developed approach has potential for different applications in these and others areas of science.

**Acknowledgement**: this research was supported by the Intramural Research Program of the NIH, NCBI.